\documentclass[journal,12pt,onecolumn,]{IEEEtran}
\usepackage{array}
\usepackage{color}
\usepackage{grffile}
\RequirePackage{afterpage}
\usepackage{epsfig}
\usepackage{here}
\usepackage{latexsym}
\usepackage{graphicx}
\usepackage{times}
\usepackage{amssymb}
\usepackage{latexsym}
\usepackage{program}
\usepackage{amssymb}
\usepackage{longtable,lscape}
\usepackage{subfigure}
\usepackage{rotating}
\usepackage{multirow}
\usepackage[linesnumbered,ruled]{algorithm2e}
\usepackage{epsfig}

{}
{}
{}

\begin{document}

\title{Vehicle Detection and Classification using LTE-CommSense}

\author{\IEEEauthorblockN{Santu Sardar\IEEEauthorrefmark{1,3},
Amit K. Mishra\IEEEauthorrefmark{2},~\IEEEmembership{Senior Member,~IEEE} and\\
Mohammed Zafar Ali Khan\IEEEauthorrefmark{3}, ~\IEEEmembership{Senior Member,~IEEE}}\\
\IEEEauthorblockA{\IEEEauthorrefmark{1}ANURAG, Defence Research \& Development Organization, India}\\
\IEEEauthorblockA{\IEEEauthorrefmark{2}Department of Electrical Engineering, University of Cape Town, South Africa}\\
\IEEEauthorblockA{\IEEEauthorrefmark{3}Department of Electrical Engineering, IIT Hyderabad, India}
\thanks{This paper is a postprint of a paper submitted to and accepted for publication in IET Radar, Sonar \& Navigation and is subject to IET Radar, Sonar \& Navigation Copyright.}

}

\maketitle

\begin{abstract}
We {\textcolor{black}{demonstrated}} a vehicle detection and classification method based on Long Term Evolution (LTE) communication infrastructure based environment sensing instrument, termed as LTE-CommSense by the authors. This technology is a novel passive sensing system which  focuses on the reference signals embedded in the sub-frames of LTE resource grid. It compares the received signal with the expected reference signal, extracts the evaluated channel state information (CSI) and analyzes it to estimate the change in the environment. For vehicle detection and subsequent classification, our setup is similar to a passive radar in forward scattering radar (FSR) mode. Instead of performing the radio frequency (RF) signals directly, we take advantage of the processing that happens in a LTE receiver user equipment (UE). We tap into the channel estimation and equalization block and extract the CSI value. CSI value reflects the property of the communication channel between communication base station (eNodeB) and UE. We use CSI values for with and vehicle and without vehicle case in outdoor open road environment.  Being a receiver only system, there is no need for any transmission and related regulations. Therefore, this system is low cost, power efficient and difficult to detect. Also, most of its processing will be done by the existing LTE communication receiver (UE). In this paper, we establish our claim by analyzing field-collected data. Live LTE downlink (DL) signal is captured using modeled LTE UE using software defined radio (SDR). The detection analysis and classification performance shows promising results and ascertains that, LTE-CommSense is capable of detection and classification of different types of vehicles in outdoor road environment. 
\end{abstract}

\section{Introduction}

{\textcolor{black}{Coexistence of telecommunication and radar systems \cite{baker,tong} to detect and track targets is an active research domain. It gains importance because devices following this principle can be low footprint and passive in nature}. There are many proposals of vehicle detection using active systems \cite{1211012,5983805,4357739}. Main principle of these systems are to transmit a signal and analyze corresponding reflected signal from the vehicle. Passive Radar or commensal Radar \cite{baker,tong,7489757,bhatta} enjoys benefit from each other without any detrimental effect on each other. The efforts of coexistence using commensal Radar principle have many real-world applications \cite{bhatta, phd_mag}. Few ideas are proposed in open literature which utilizes commensal radar principle for vehicle detection and classification \cite{cher,raja}. In \cite{raja}, the incident spectrum was analyzed using their custom RF receiver model. Later cluster analysis of the principal components of the spectrum was performed.  

One innovative way of utilizing the telecommunication system is to collect information from the communication receiver modules which are responsible for reducing the channel imperfections. Channel estimation is performed to estimate the channel conditions by observing effect of channel on a known data pattern. These known data patterns are embedded within the payload at pre-defined locations. They are called reference symbols or pilots. After that, the channel equalization block is present to nullify the effect of estimated channel imperfections. Therefore, the output of channel estimation and equalization block contains channel information. The scattering, fading, power reduction with distance of the communication link are collectively known as channel state information (CSI). This CSI value is exploited by proposed CommSense scheme \cite{africon, akm_patent}. In the past two years, a good amount of work has been done regarding the feasibility of a GSM (Global System for Mobile telecommunication) based CommSense system \cite{7489757,bhatta}. Suitability of LTE based CommSense system is presented in \cite{africon,phd_mag}. LTE networks are widespread and follows Orthogonal frequency division multiplexing (OFDM) Principle.  Wideband nature of OFDM can overcome fading and multipath problems  and have low interception probability as well \cite{Huang2015LowPO}. Therefore we can assume that those information gets reflected in the evaluated CSI. Using frequency diversity, more information about the environment change of interest may be obtained. Therefore, LTE is selected for CommSense based vehicle detection and classification application. This proposed idea for vehicle detection and classification has the following novelty and advantages.

\begin{itemize}
\item Use of CSI extracted from LTE DL signal in Commensal radar setup for vehicle detection and classification is a novel effort.
\item A part of the processing gets done by the LTE communication receiver itself. This reduces the custom computation efforts and additional processing modules and infrastructure.
\item {\textcolor{black}{As the communication base station will work as the transmitter, there is no requirement of a dedicated transmitter to be a part of our proposed sensing instrument}}. Therefore, the system complexity to develop such instrument is less and it doesn't require to adhere to any transmission regulations. This ensures quick and easy installation for practical applications. 
\item Without any transmission by this instrument and low footprint makes the system difficult to detect. Therefore, the proposed system is suitable for covert defence operations.
\item Reuse of outputs of communication receiver modules and no requirement of signal transmission makes the system low power and less costly. This makes the system suitable for battery operated field applications.
\end{itemize}

The following results are shown in this paper: 

\begin{enumerate}
\item First, the feasibility of using LTE CommSense instrument for vehicle detection is investigated. LTE downlink (DL) signal was captured using CommSense system implemented using an Universal Software Radio Peripheral (USRP) SDR platform. CSI values are extracted from the captured DL data for the case of empty road environment and with vehicle case in five different outdoor open road environment. Principal component analysis (PCA) are performed on the CSI values and optimal number of principal components are selected by considering the energy contribution and cluster analysis. After that, threshold based detection method was devised and vehicle detection performance was evaluated.
\item In the next stage, classification between empty road  and three different types of vehicle namely a two wheeler (Honda Acticva), a sedan car (Hyundai I20) and an sport utility vehicle i.e. SUV (Ford EcoSport Titanium) was exercised. To evaluate the classification performance, classification accuracy, confusion matrix, false acceptance rate (FAR) and false rejection rate (FRR) were evaluated. 
\end{enumerate} 

The paper is organized as follows. Section 2 describes the basic structure of LTE CommSense, system design for the application scenario of vehicle detection and classification. Section 3 provides the experiments performed, their objectives. Results and data analysis for vehicle detection and classification are detailed in Section 4. The conclusion of this work and future scopes of improvements are provided in Section 5.

\section{LTE-CommSense System}

General block diagram and description of LTE-CommSense are elaborated in \cite{africon,phd_mag}. Figure \ref{block} provides a simple conceptual system described in \cite{africon,phd_mag} using single UE which is implemented and utilized in this work of vehicle detection and classification. 

\begin{figure}[!h]
\centering
\psfig{file=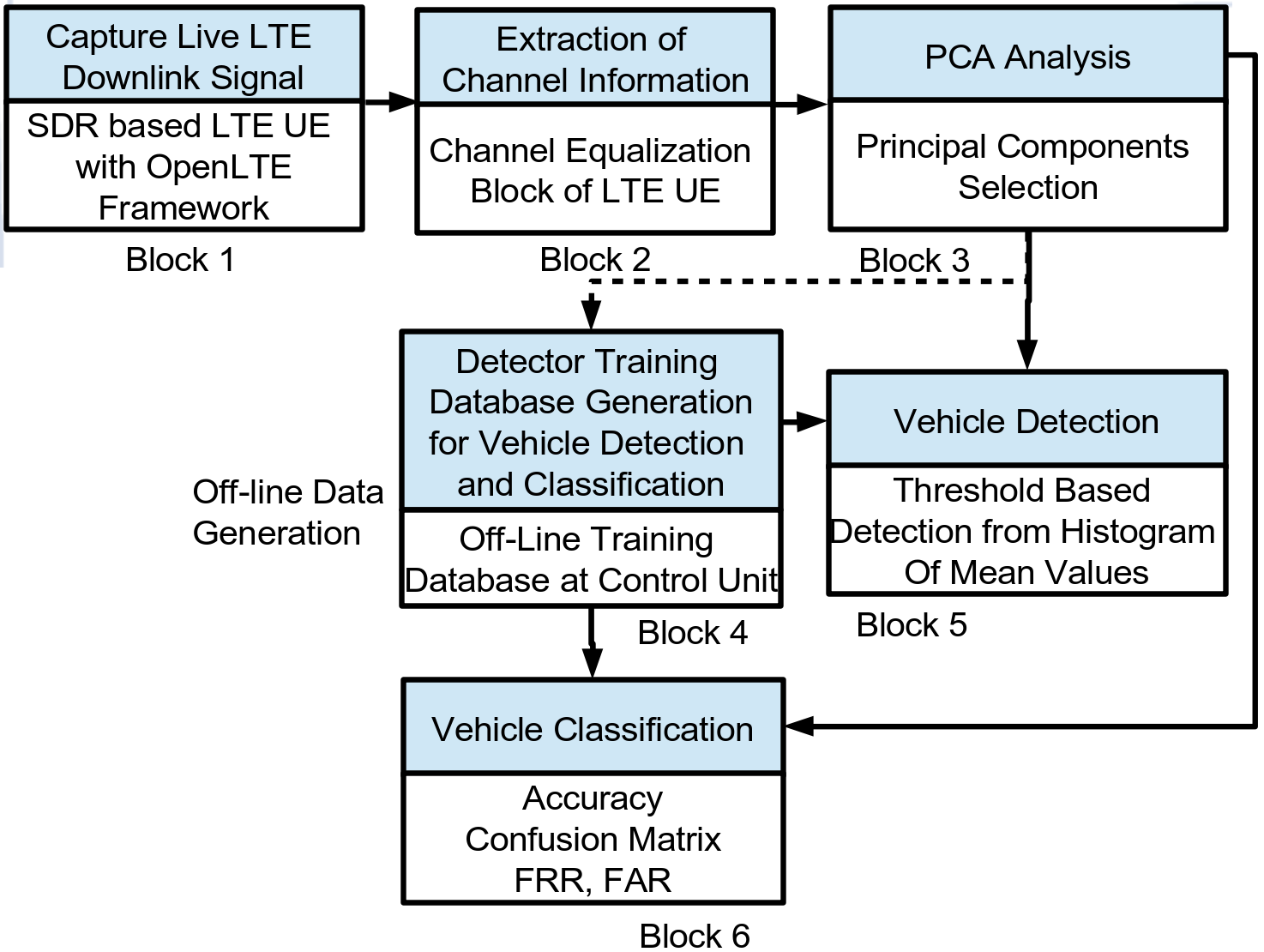,width=0.5\textwidth}
    \caption{LTE-CommSense instrumentation system block diagram using single UE for Vehicle Detection and Classification.}
    \label{block}
\end{figure}

There are two parts of processing in the system: one is offline and the other one is online. A brief description of individual blocks are as follows.

\begin{itemize}
\item Block 1: This block first detects and then captures LTE DL signal from the `eNodeB' using the USRP N200 SDR platform. This SDR is configured as an LTE UE using OpenLTE framework. This block is used both in offline and online modes. 
\item Block 2: In the LTE UE receiver work-flow, the channel equalization block extracts the channel information from the channel affected call specific reference symbols (cellRS) in the form of channel state information (CSI). The objective of this block is to tap in to the channel equalization block of the UE to capture the extracted channel information. This block is also used both in offline and online mode. 
\item Block 3: It is used in online and offline mode. For vehicle detection and classification application using LTE-CommSense, we perform principal component analysis (PCA) of the extracted channel state information (CSI). The eigen values are arranged in descending order and their energy contribution to select optimal number of principal components for our analysis.
\item Block 4: We have to generate the database to train the LTE-CommSense using the generated principal components of CSI values. Data corresponding to multiple capture corresponding to five different road background and three different vehicles are used to generate the training database and maintained in the control unit. The control unit tunes LTE-CommSense for targeted application.
\item Block 5: This is the LTE-CommSense detector and used in both modes of operation. It accepts principal components of a CSI corresponding to a DL signal capture of current channel condition. The objective of the detector is to give decision of regarding a vehicle is present or not. 
\item Block 6:  {\textcolor{black}{This is the vehicle classifier using LTE-CommSense principle and is used in both the modes of operation. First, it works in offline mode and gets trained by the vehicle database provided by the control unit in `Block 4'. After that in online mode, it accepts principal components of a CSI corresponding to a DL signal capture. The classifier provides us the classification accuracy. In the feasibility analysis presented in this paper, we evaluate classification accuracy, confusion matrix, false acceptance rate (FAR) and false rejection rate (FRR) to carry out the performance analysis of the proposed classifier and decide its classification capability. The Naive Nearest Neighbour classifier is selected in this feasibility analysis.}}
\end{itemize}

Previous works on CommSense have shown promising results using both GSM and LTE telecommunication standards (\cite{7489757,bhatta,africon,phd_mag}). Our proposed hypothesis in this work is that, using extracted CSI values of LTE downlink signal, vehicle can be detected and multiple vehicle can be classified. CSI can be termed as known channel properties of a communication link. CSI provides information regarding how communication downlink signal traverses from the base station (eNodeB in case of LTE). It represents the combined effect of the channel imperfections i.e. decay of signal power with respect to distance, channel fading and scattering. In communication CSI helps to adjust the transmission to the present condition of the channel. It is estimated in the receiver (UE in case of LTE) and sent back to transmitter. In LTE communication, Cell specific reference symbols (CellRS) are used to evaluate the CSI values. Mathematically, let the reference symbols be $(P = p_1,p_2,...,p_N)$ and CSI be $H$. Corresponding received signals at the UE can be represented as:

\begin{equation}
y_i = H*p_i+n_i; i = 1:N.
\end{equation}

Representing in vector form for all the training signal, we obtain,

\begin{equation}
Y = [y_1,y_2,...,y_N] = H*P+N; N = [n_1,n_2,...,n_N];
\end{equation}

{\textcolor{black}{Here, $H$ is the statistical channel model of the channel in between communication transmitter and the receiver and $N$ is the model of noise in the channel}}. $H$ is modeled using circular symmetric complex Normal distribution i.e. $n_i \thicksim cN(0,\sigma)$; where $\sigma$ is the standard deviation of the Normal distribution. Corresponding Least Square (LS) estimate of the CSI is given as:

\begin{equation}
H_{LS} = YP^H(PP^H)^{-1}
\end{equation}

If channel distributions and noise distributions are known a prior, then estimation error will be reduced. For Minimum Mean Square Estimation (MMSE) in Rayleigh fading channel, if we know that $Vec(H) \thicksim cN(0,R)$ and $Vec(N) \thicksim cN(0,\sigma)$, where $R$ is the channel covariance matrix, then,
$$ Vec(H_{MMSE}) = R^{-1}+... $$
\begin{equation}
(P^T\otimes I)^HS^{-1}(P^T\otimes I)^{-1}(P^T\otimes I)^HS^{-1}vec(Y)
\end{equation}

Here, $\otimes$ denotes the Kronecker product \cite{7410692}. In \cite{africon}, we have established that, MMSE estimator is more suitable candidate for CommSense based applications. Accordingly it is selected here.

In case of LTE communication, CSI values gets continuously evaluated using embedded $CellRS$ symbols in the LTE subframes. As these CSI values contains the channel information in between eNodeB and UE, we propose to use this to detect and classify vehicles on road. To perform this, we devise the following experimental setup.

\section{Experimental Setup} 

The objective of this experiment is two-fold. First we will evaluate the feasibility of detecting a vehicle in outdoor environment. If detected, we try to see the distinguishing capability of our proposed system. For that, we consider three different types of vehicles viz. a two wheeler (Honda Acticva), a sedan car (Hyundai I20) and an sport utility vehicle i.e. SUV (Ford EcoSport Titanium) as shown in Figure \ref{cars}. A brief structural description of the considered vehicles are provided in Table \ref{car_dim}.

\vspace{-9pt}
\begin{table}[!h]
\caption{Structural Descriptions of the Vehicles Considered for Evaluation of Vehicle Classification Performance of LTE-CommSense.}
\begin{center}
\begin{tabular}{|m{33mm}|m{33mm}|m{10mm}|}\hline
Vehicle Name & Dimension (Length*Width*Height) & Weight     		\tabularnewline		\hline
Honda Acticva 			& 1761*710*1158 ($mm^3$)    	&	109 Kg	   	 \tabularnewline	\hline
Hyundai I20 				& 3985*1734*1505 ($mm^3$)	&	 1066Kg 	 \tabularnewline	\hline
Ford EcoSport Titanium 	& 3998*1765*1708 ($mm^3$)	&	 1320Kg          \tabularnewline	\hline
\end{tabular}
\end{center}
\label{car_dim}
\end{table}
\vspace{-15pt}

\begin{figure}[!h]
\centering
\subfigure[]{\psfig{file=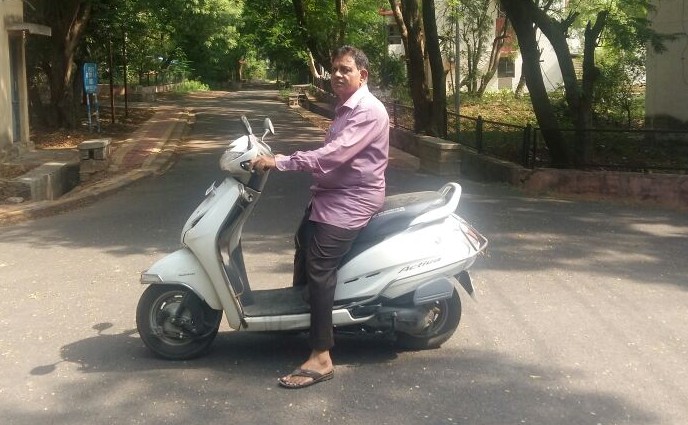,width=.32\textwidth}}  
\subfigure[]{\psfig{file=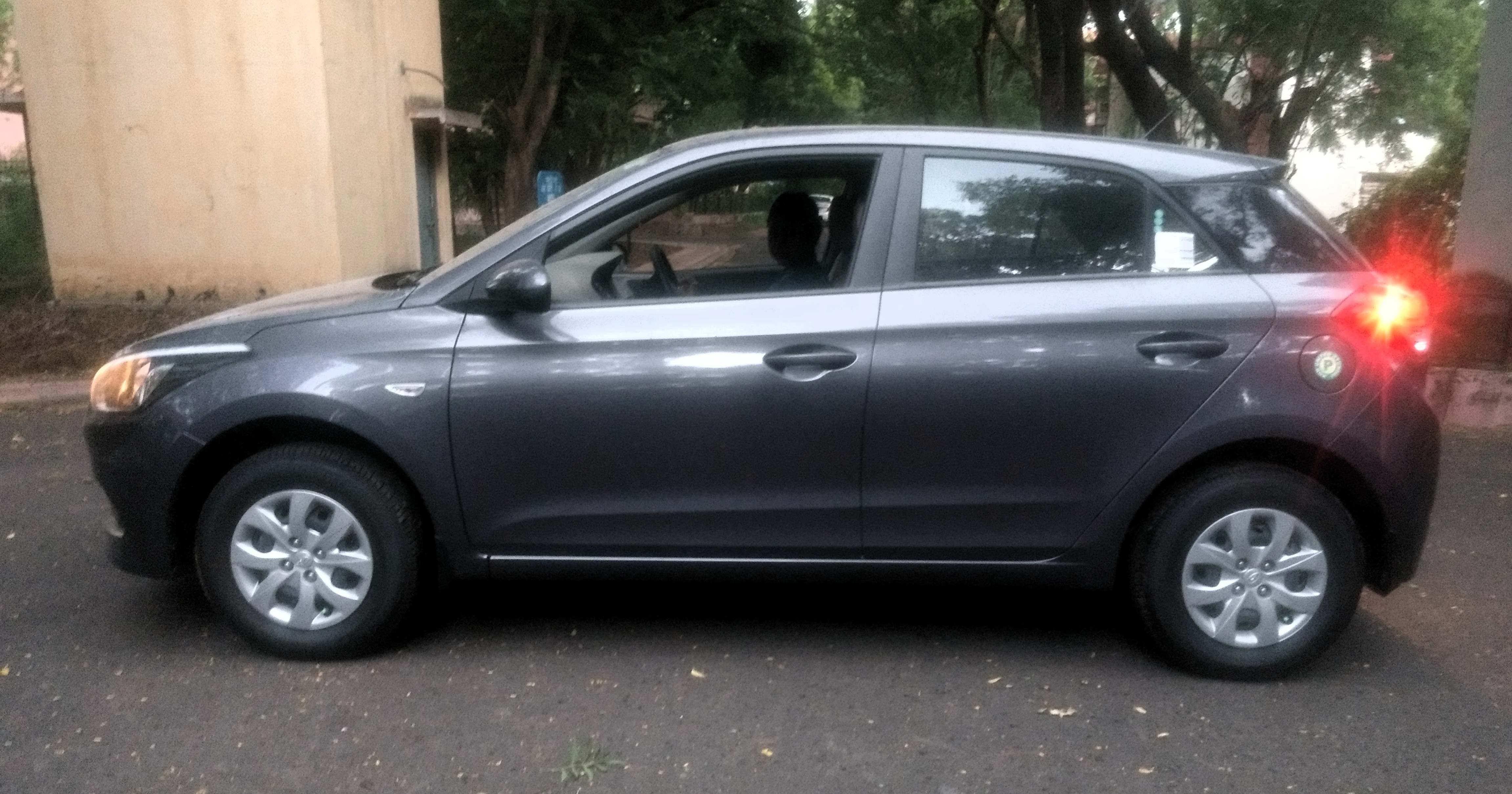,width=.32\textwidth}}  
\subfigure[]{\psfig{file=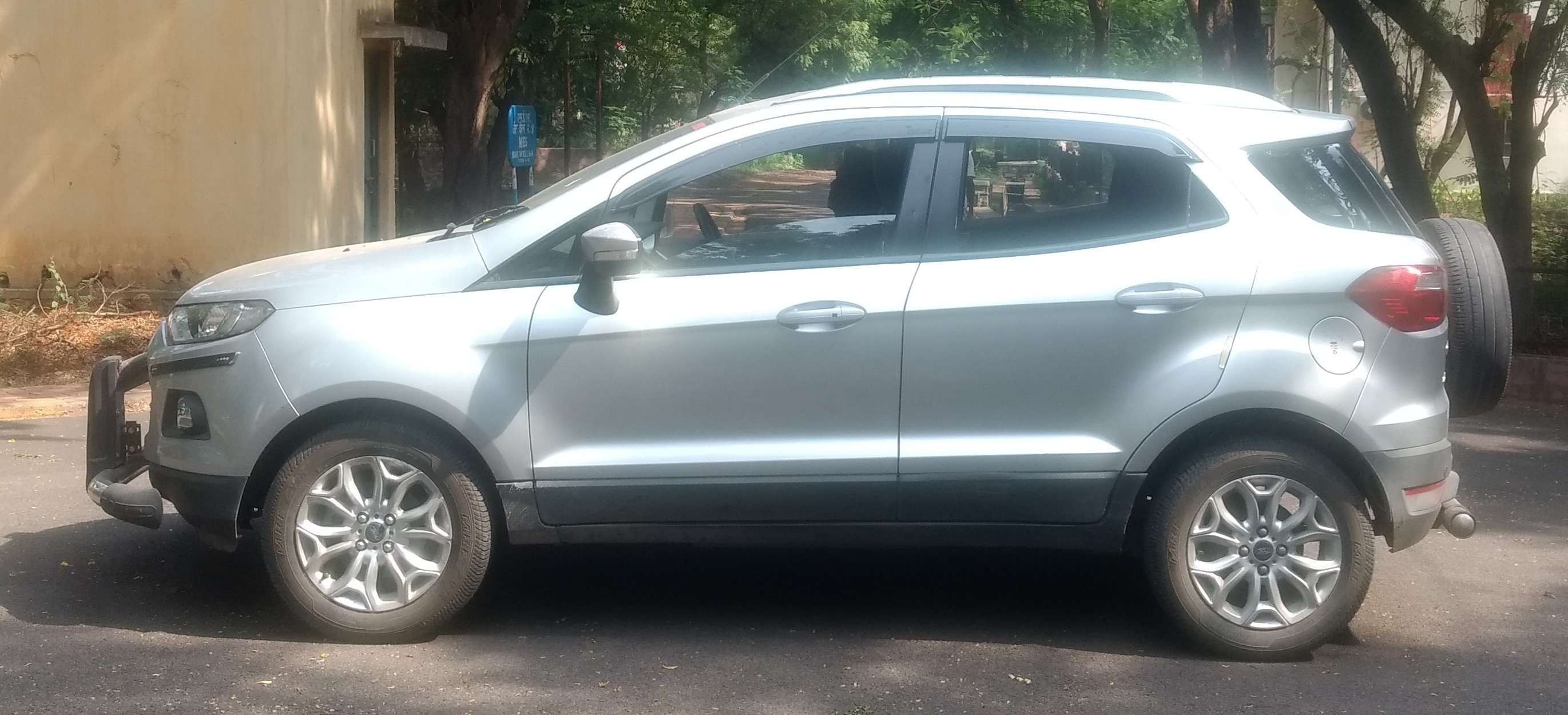,width=0.32\textwidth}} 
\caption{Different Vehicles used in the Experiment: (a) 2 Wheeler: Honda Activa, (b) Sedan Car: Hyundai I20 (c) SUV car: Ford EcoSport Titanium.}
\label{cars}
\end{figure}

To obtain live LTE downlink signal, SDR based data capturing system was developed. USRP N200 SDR platform \cite{usrp} and RFX2400 RF daughter card \cite{rfx2400} along with VERT2450 antenna \cite{antenna} was used to work at LTE passband operating frequency range. A GNURadio based open source model of the 3GPP LTE receiver on OpenLTE \cite{openlte} was utilized to control the SDR platform. The capturing SDR device N200 along with OpenLTE implemented on it and interfaced host computer are shown in Figure \ref{usrpn200}(a). The internal block diagram of USRP SDR platform is shown in Figure \ref{usrpn200}(b). The SDR unit was connected to the host computer via Ethernet. The OpenLTE framework is implemented on the Host computer. The commands for searching for live LTE DL signal, capturing detected LTE DL data on the host computer for a specified time duration were also input to the SDR platform via host PC using the OpenLTE framework. The LTE UE receiver works at $Band-40$ (2300 MHz to 2400 MHz frequency band) in time division duplexing (TDD)  topology. {\textcolor{black}{The experiments performed for this feasibility analysis exploits a $15 MHz$ bandwidth (BW) of LTE downlink signal}}. Figure \ref{usrpn200}(c) shows the SDR platform relative placements of the vehicle for this experiment. The vehicle is situated on a road in outdoor environment. {\textcolor{black}{In this figure, the commercial LTE base station, $eNodeB$ is located $1.35$ $km$ away from our object of interest i.e. vehicle/car. The user equipment (UE) modeled using the SDR platform is situated at a distance of $2$ $meters$ meter from the car. This distance is considered for the situation where the road is $4$ $meters$ meter wide, the vehicle/car is at the middle of the road and our sensing instrument is stationed at the side of the road.}}

For each of the vehicle, LTE downlink signal is captured for five different outdoor road environment. For each capture, 1000 CSI values were extracted from the LTE receiver channel equalization block.

\begin{figure}[!h]
\centering
\subfigure[]{\psfig{file=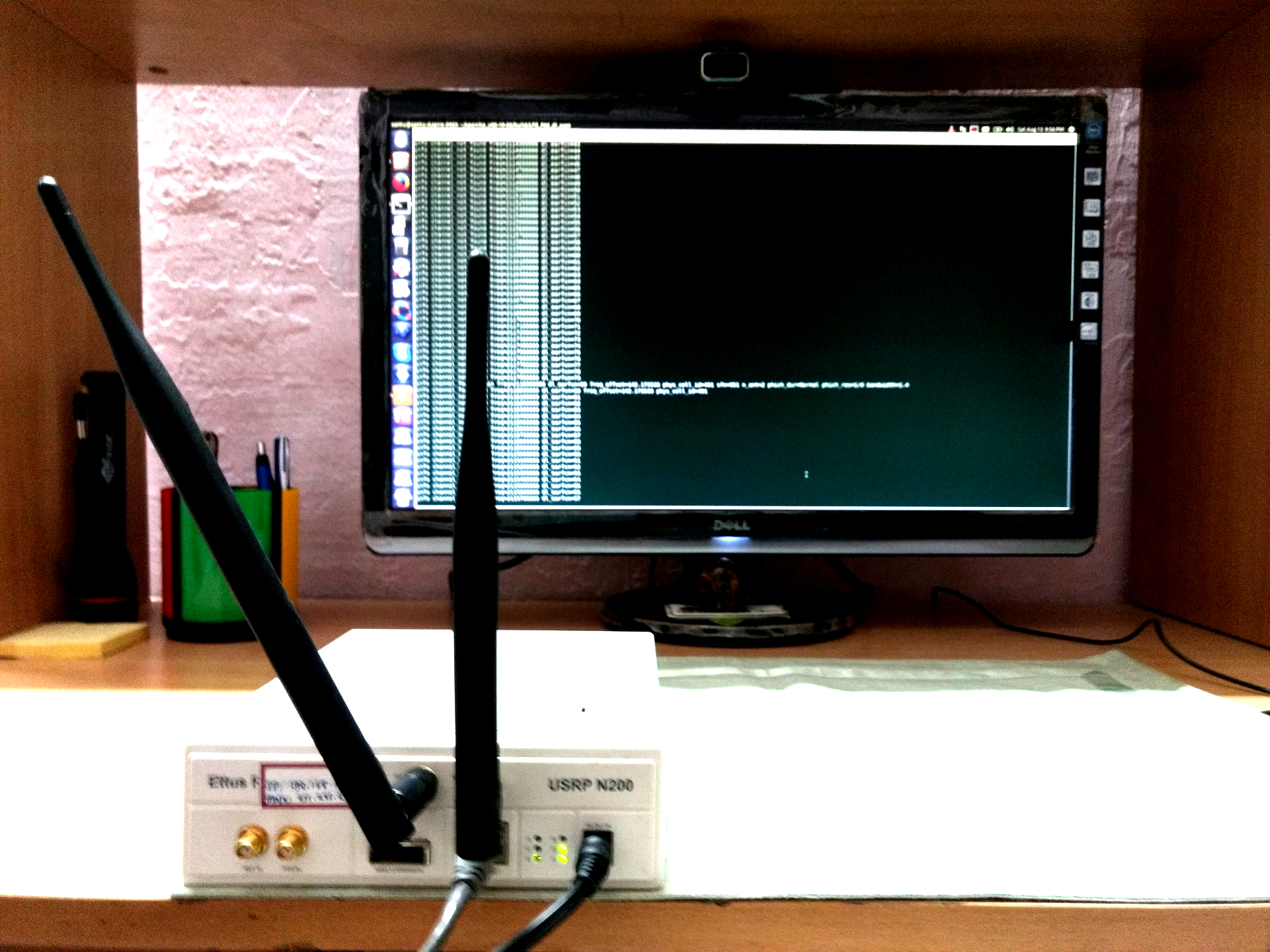,width=0.45\textwidth}}  
\subfigure[]{\psfig{file=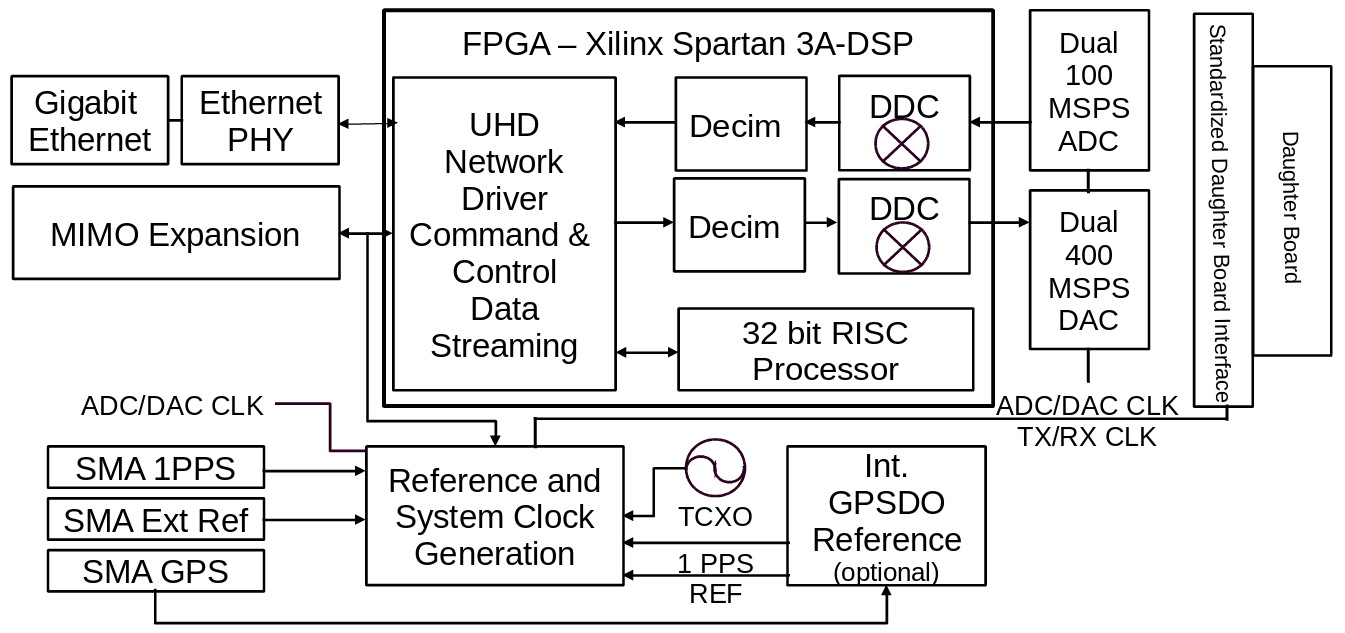,width=0.45\textwidth}}  
\subfigure[]{\psfig{file=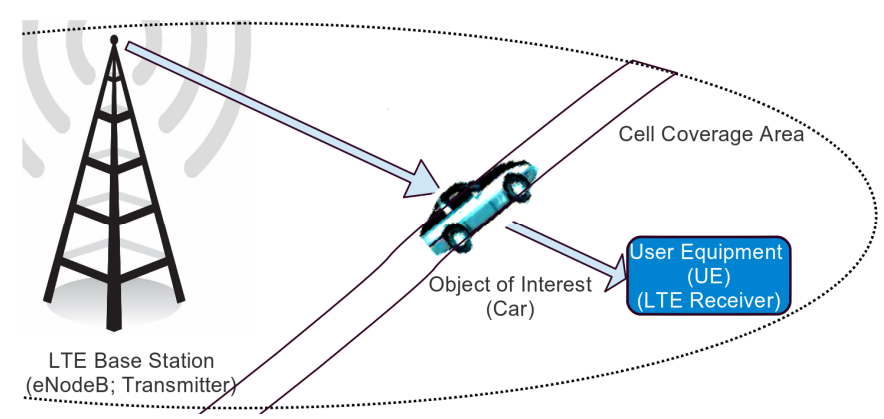,width=0.45\textwidth}} 
\caption{(a) USRP N200 SDR platform (front) working as an LTE UE to Capture Live LTE Signal and Host PC screen showing detected LTE channel using OpenLTE, (b) USRP N200 Internal Block Diagram showing individual components, (c) Position of Receiver (LTE UE), Vehicle and Transmitter (eNodeB) in Outdoor Environment.}
\label{usrpn200}
\end{figure}

\section{Results and Analysis}

Vehicle detection is exercised first. After detection, classification is performed to evaluate the distinguishing capability of proposed scheme between different types of vehicles.

\subsection{Vehicle Detection}

CSI values evaluated for five different background situations. For each background, CSIs are evaluated for only background and with vehicle case. The vehicle considered in this case is sedan car (Hyundai I20). For each of the cases explained above 1000 CSI values were evaluated from the captured DL data. 

\subsubsection{CSI Analysis}

Figure \ref{csi} Shows CSI for with and without vehicle case and their differences for each of the five different backgrounds We can observe a clear and consistent CSI difference for each case.

\begin{figure}[!h]
\centering
\psfig{file=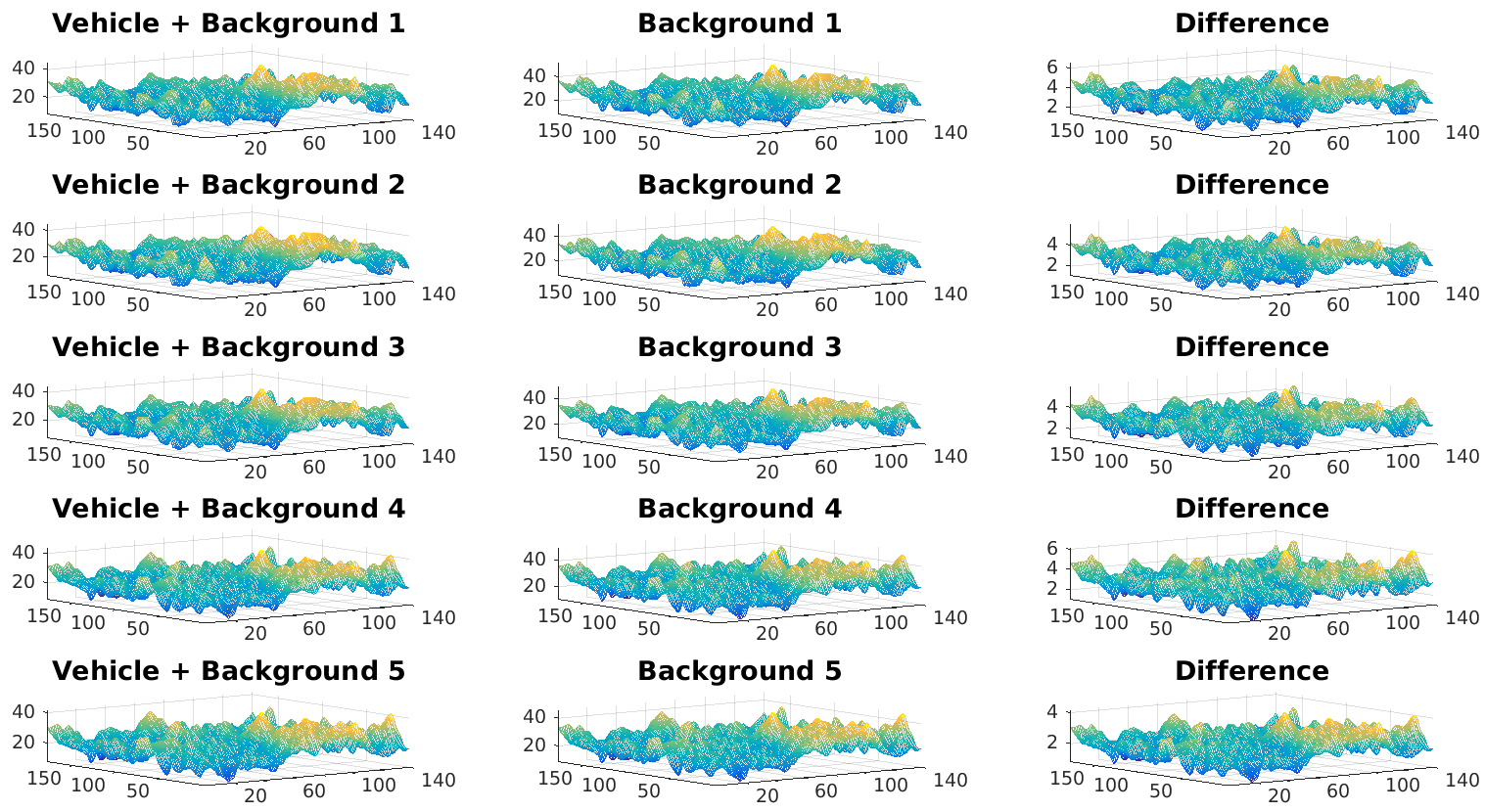,width=0.5\textwidth} 
    \caption{Evaluated CSI Information for With and Without Vehicle Case in Five Different Outdoor Environment.}
    \label{csi}
\end{figure}

\subsubsection{PCA Analysis}

Figure \ref{pca_scatter}(a) shows the eigenvalues arranged in decreasing order and their energy contribution. It can be observed that, less first three of eigenvalues contribute to 99.9\% cumulative energy. {\textcolor{black}{As the first eigenvalue contributes to 99.79\% of the total cumulative energy, it can be stated that the largest component have a useful discrimination capability.}}

Figure \ref{pca_scatter}(b) and \ref{pca_scatter}(c) shows the two dimensional (2D) and three dimensional (3D) scatterplot corresponding to highest two and highest three principal components respectively. {\textcolor{black}{In case of 2D scatterplot, X-axis is the direction of the principal eigen-vector corresponding to  largest eigenvalue and Y-axis is the direction of the eigen-vector corresponding to second largest eigenvalue. Additionally in case of 3D scatterplot, Z-axis is direction of the eigen-vector corresponding to third largest eigenvalue. Projection of the plotted points to the respective axis provides corresponding principal components. The largest principal component corresponds to the largest eigenvalue. Therefore, the  largest principal component has maximum discriminating power. This is more evident from Figure \ref{pca_scatter}(b). It can be observed that, the projection of the points onto the X-axis can be separated with minor overlaps using a single selected threshold.}}

{\textcolor{black}{In the figure, the blue points corresponds to the case of data captures in presence of vehicle on the road and black points are for the case of without vehicle}}. It can be easily observed that, two distinct clusters with minor overlap between them are visible in the 2D and 3D scatterplots.

\begin{figure}[h]
\centering
\subfigure[]{\psfig{file=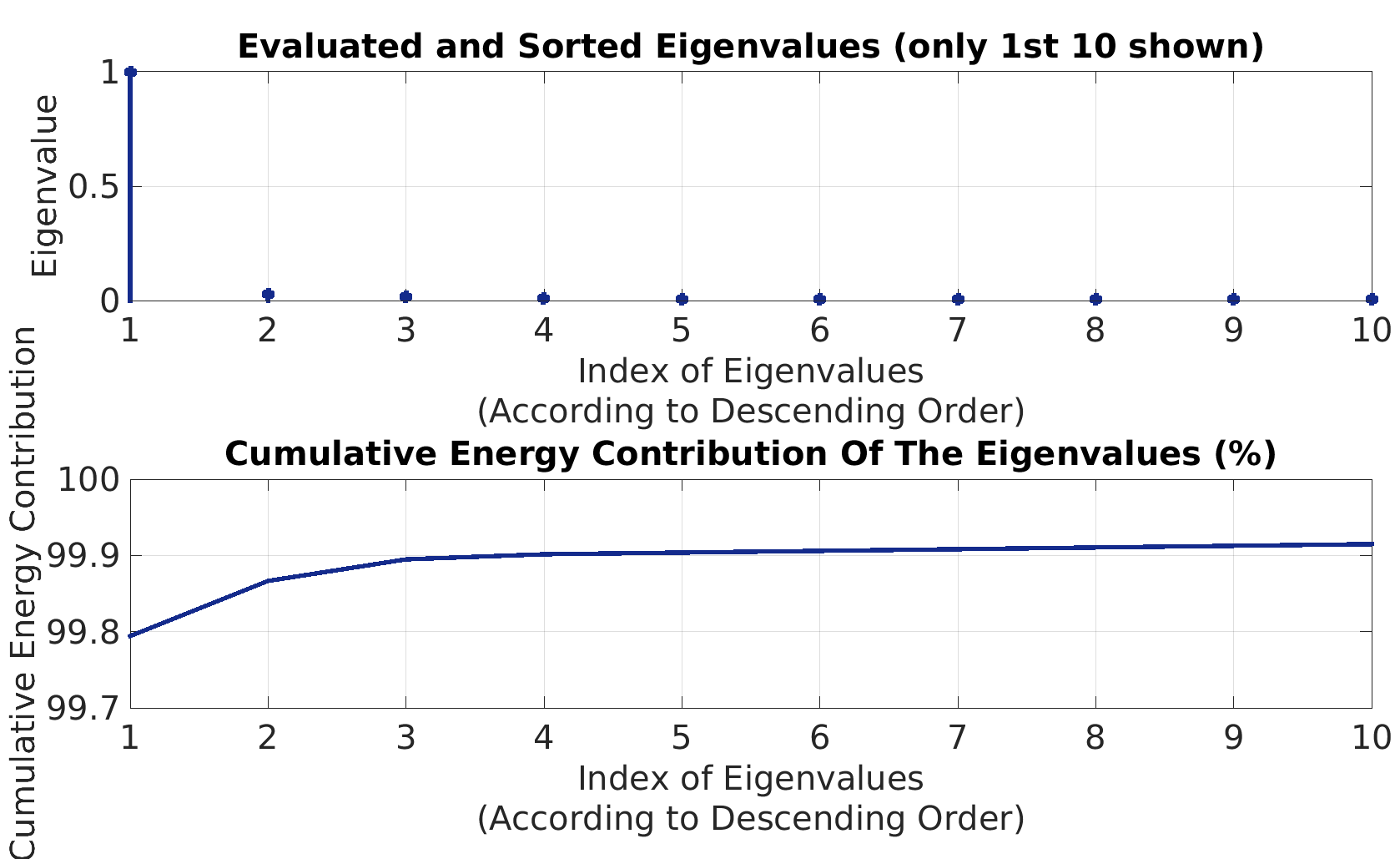,width=.45\textwidth}}  
\subfigure[]{\psfig{file=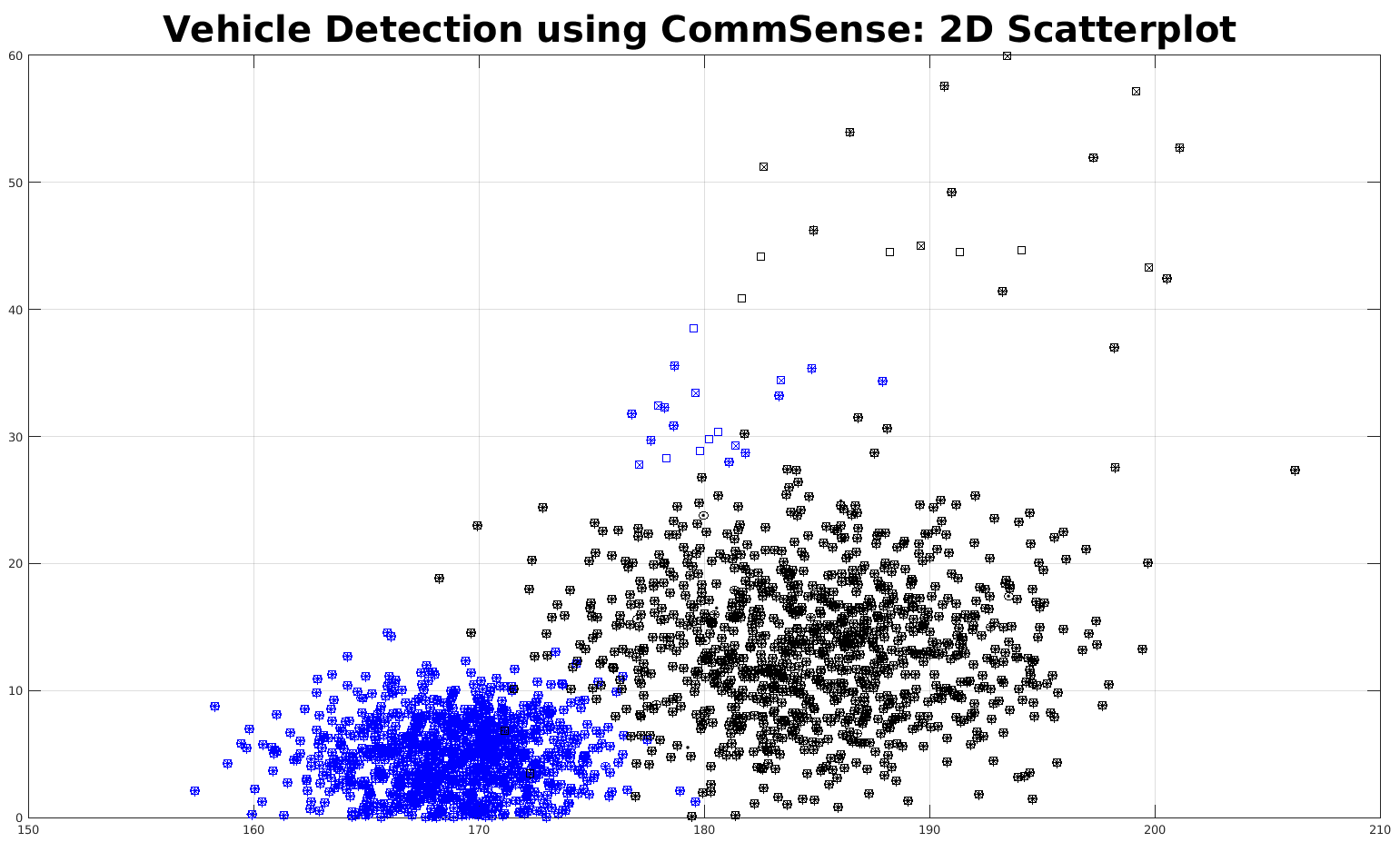,width=.45\textwidth}}  
\subfigure[]{\psfig{file=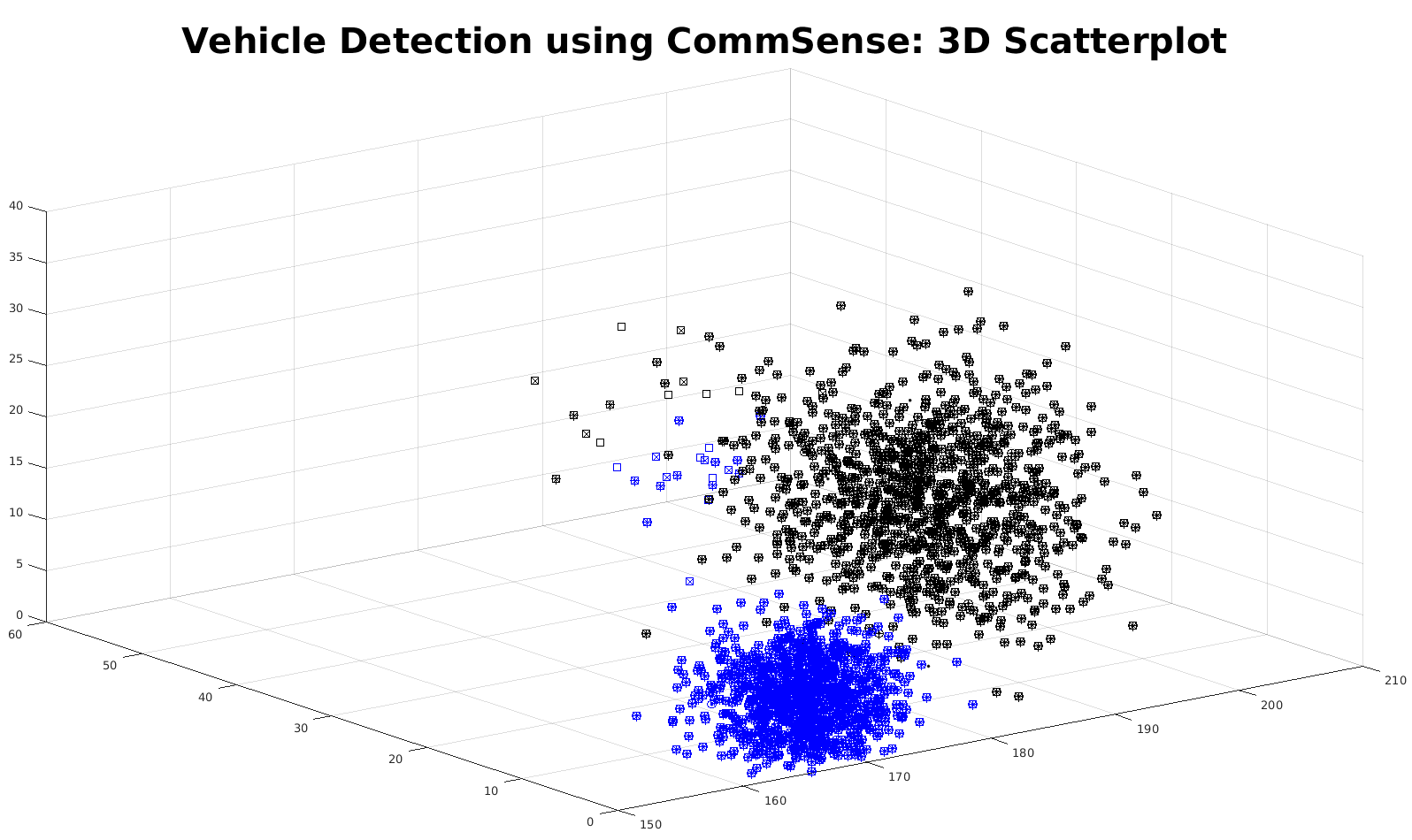,width=.45\textwidth}}  
\caption{ (a) Highest 10 Eigenvalues and Their Corresponding Energy Contribution, (b) Scatterplot for Highest Two Principal Components, (c) Scatterplot for Highest Three Principal Components.}
\label{pca_scatter}
\end{figure}

\subsubsection{Threshold Selection and Detection Performance}

Figure \ref{histogram} shows the histogram of mean of considered number of principal components for with (blue) and without vehicle (orange) case. Each subfigure is for a specific number of PCs considered and mentioned in figure. It can be observed that, mean values for both the cases forms seperate regions but overlap is there. 

Figure \ref{err} shows the selection of threshold and corresponding error percentages for a number of principal components considered. This is performed for {\textcolor{black}{different number(s) of principal components}}, shown in each subfigure. We have considered a simple detection rule: if mean of considered principal components is less than the threshold then vehicle is present; else  vehicle is absent. It is observed that for very less number of principal components, minimum error percentage is more. This simple threshold based method can achieve as low as $10\%$ detection error as shown in Figure \ref{err}.

\begin{figure}[!h]
\centering
\psfig{file=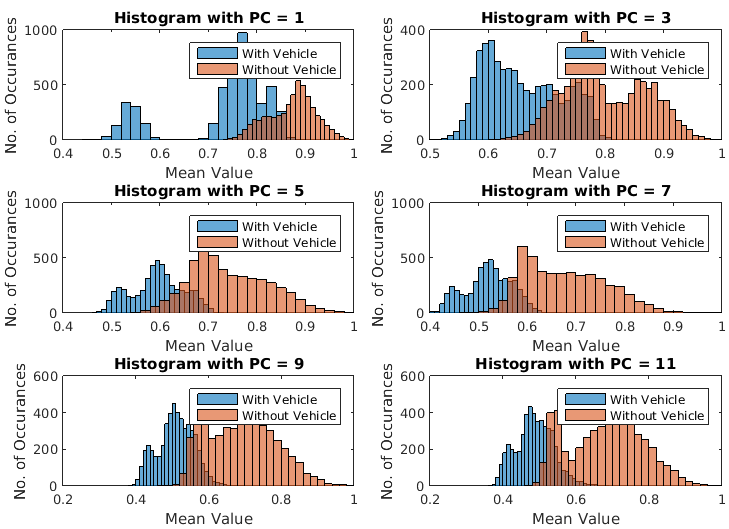,width=0.5\textwidth} 
    \caption{Histogram of the Mean Values of Principal Components for With and Without Vehicle Case Considering {\textcolor{black}{Different Number(s) of Principal Components}}.}
    \label{histogram}
\end{figure}
\begin{figure}[!h]
\centering
\psfig{file=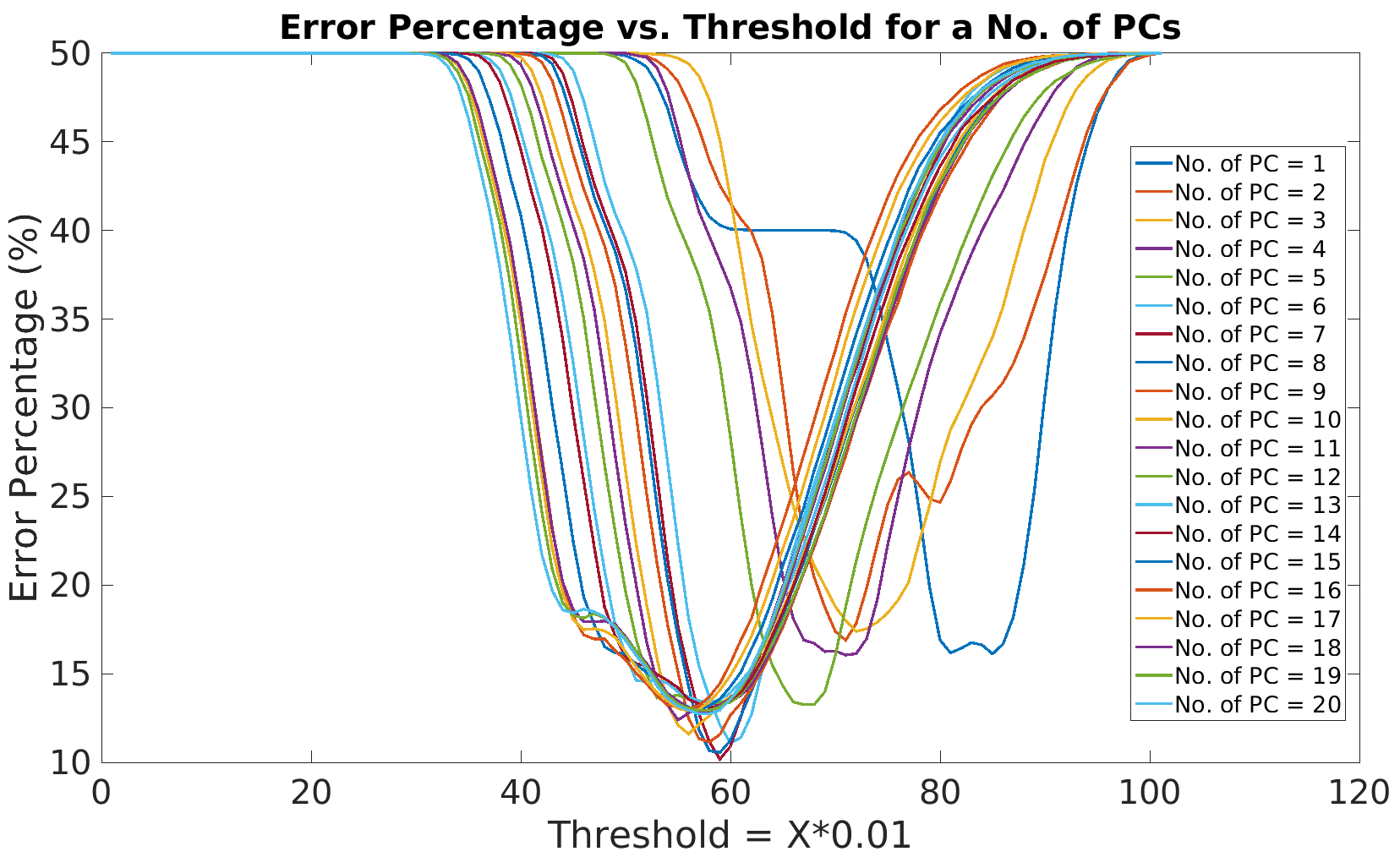,width=0.5\textwidth} 
    \caption{Threshold Selection: Error Performance with respect to Threshold Values for {\textcolor{black}{Different number(s) of Principal Components}}.}
    \label{err}
\end{figure}

Following can be concluded from this vehicle detection feasibility analysis.

\begin{itemize}
\item The CSI plots are consistent for with vehicle and without vehicle cases as shown in Figure \ref{csi}. 
\item CSI pattern for with vehicle case are different than without vehicle case. Their differences are also consistent as shown in the third column of the Figure \ref{csi}. 
\item The scatterplots in 2D and 3D shows distinct clusters for with and without vehicle cases.
\item We observe that, using this simple detection rule, we are able to achieve more than $90\%$ detection accuracy. This proves that vehicle detection is feasible using LTE-CommSense infrastructure. 
\end{itemize}

\subsection{Vehicle Classification}

After we have established that, LTE-CommSense can be successfully used for detection of vehicle on road, next we investigate whether it can differentiate between different types of vehicle or not. {\textcolor{black}{The Nearest Neighbour classifier is used here for classification. The Nearest Neighbour classifier is the simplest classifier. Every test data is compared with every labelled trained data to predict its label or class. For this comparison, $L1$ norm based distance measure is selected. To calculate $L1$ norm, we compare two data by adding all the absolute difference of corresponding data-points. In future, this can be replaced with better classifier, if needed.}}

\subsubsection{PCA Analysis}

Figure \ref{pca_scatter1}(a) shows the eigenvalues arranged in decreasing order and their corresponding energy contribution. We can see that, less first three of eigenvalues contribute to 99.9\% cumulative energy as was already seen during detection process also. 

Figure \ref{pca_scatter1}(b) and \ref{pca_scatter1}(c) shows the two dimensional (2D) and three dimensional (3D) scatterplot corresponding to highest two and highest three principal components respectively for three vehicles and the background. {\textcolor{black}{Here, the Black points corresponds to the case of data captures for the road and backgrounds only, Red points are for Honda Activa, Blue points corresponds to Hyundai I20 vehicle and Green points are for the case of Ford EcoSport Titanium}}. In 2D and 3D scatterplots, it can be observed that, Four distinct clusters are visible. Also, the clusters corresponding to road background and Honda Activa are nearer to each other and so is the clusters of Hyundai I20 and Ford EcoSport Titanium. This means that, distinguishing between road background and Honda Activa is more difficult compared to distinguishing between road background and Hyundai I20 or Ford EcoSport Titanium. Similarly, distinguishing between Hyundai I20 and Ford EcoSport Titanium is more difficult compared to distinguishing between Ford EcoSport Titanium and road background or Honda Activa. In the next stage, we will numerically analyze this distinguishing ability property in terms of Classification performance, confusion matrix, false acceptance rate (FAR) and false rejection rate (FRR). 

\begin{figure}[h]
\centering
\subfigure[]{\psfig{file=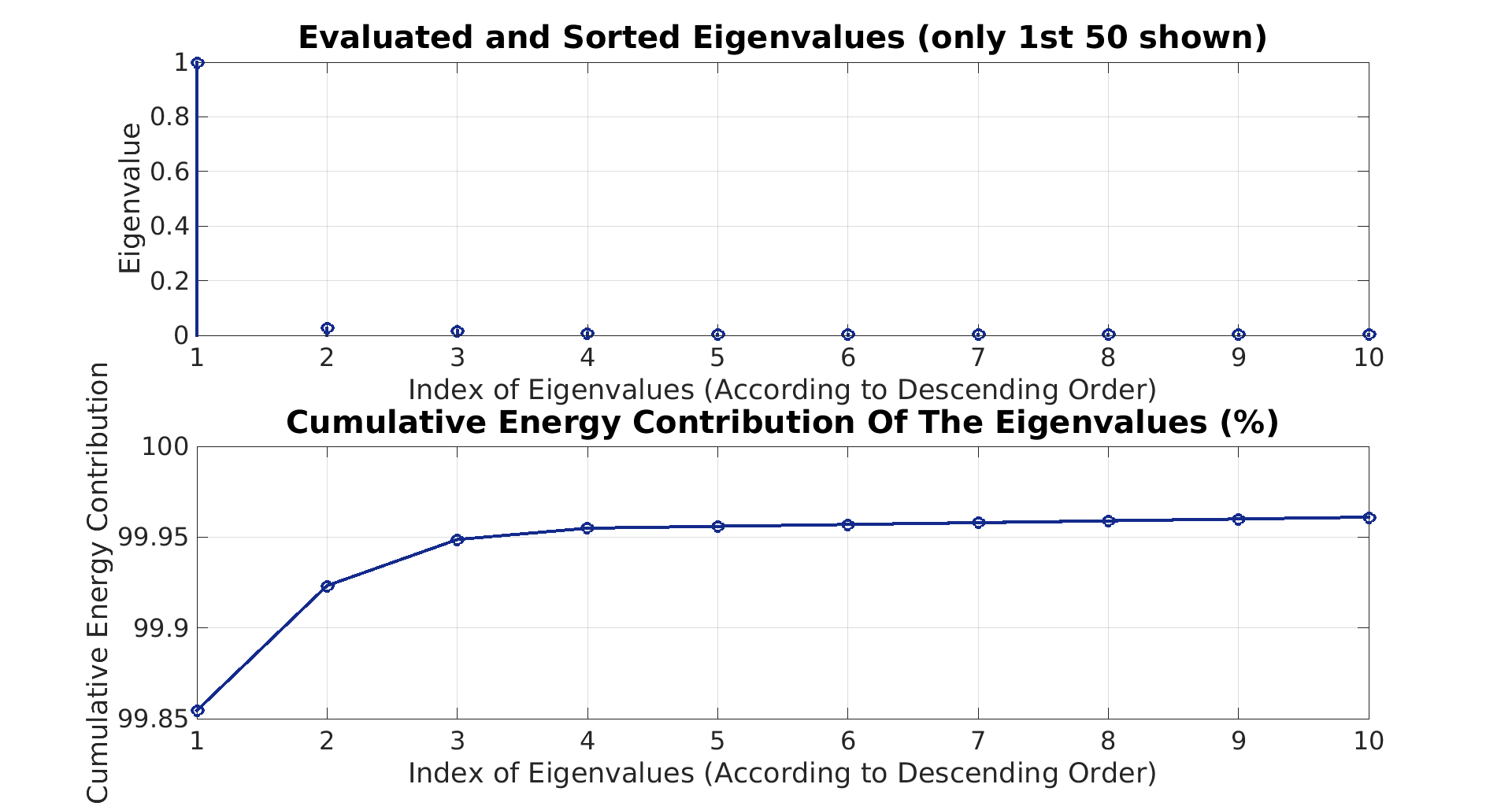,width=.45\textwidth}}  
\subfigure[]{\psfig{file=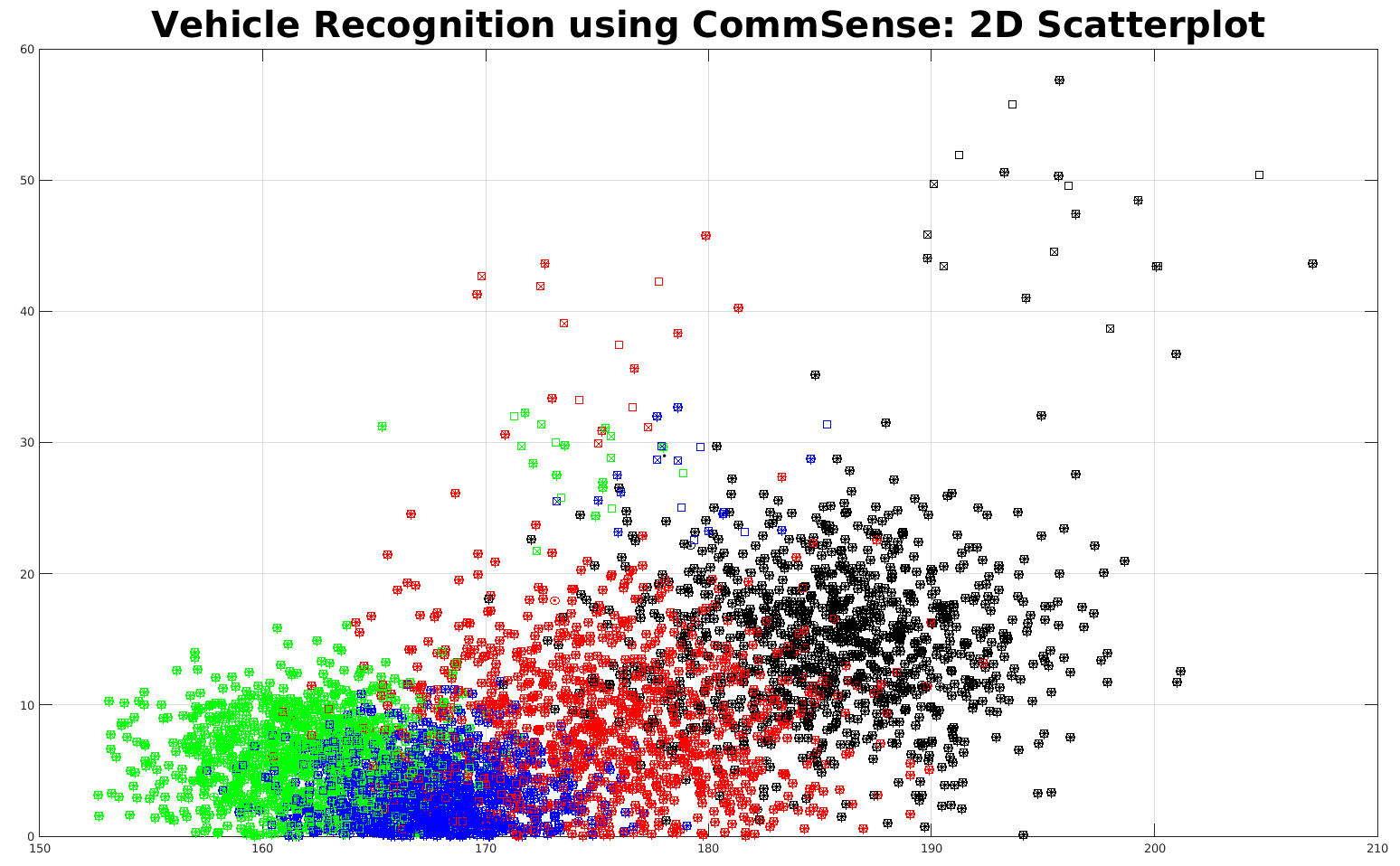,width=.45\textwidth}}  
\subfigure[]{\psfig{file=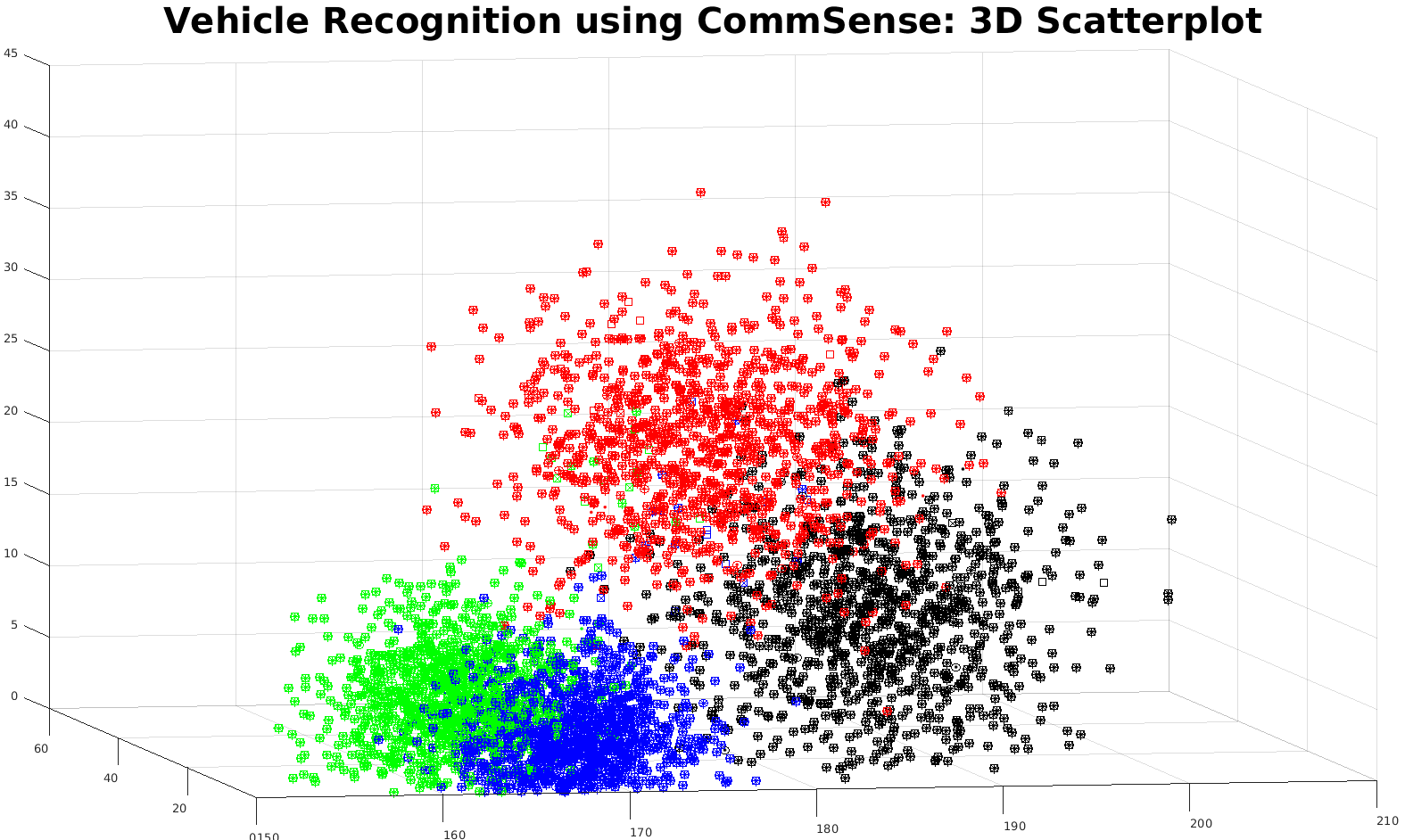,width=.45\textwidth}}  
\caption{ PCA Analysis for Vehicle Classification: (a) Highest Ten Eigenvalues and Their Corresponding Energy Contribution, (b) Scatterplot for Highest two Principal Components for Three Different Vehicles and Outdoor Background (Black: Outdoor Background, Red: Bike, Blue: Sedan, Green: SUV), (c) Scatterplot for Highest three Principal Components  for Three Different Vehicles and Outdoor Background (Black: Outdoor Background, Red: Bike, Blue: Sedan, Green: SUV).}
\label{pca_scatter1}
\end{figure}

\subsubsection{Classification Performance}

The accuracy of a classifier (\cite{1304000,78300}) denotes the extent with which the system provides correct classification output i.e. the system's ability to distinguish between occurrence of different types of changes in the environment. False Rejection Rate (FRR) or false negative rate (Type I Error) and false acceptance rate (FAR) or false positive rate (Type II Error) \cite{farfrr} are another two  performance parameters for classification. 

FAR is fraction of the falsely accepted patterns divided by the number of all impostor patterns. The fraction of the number of rejected client patterns divided by the total number of client patterns is called false rejection rate (FRR). The frequency at which false acceptance errors are made can be denoted as false acceptance rate (FAR). If $N_{FA}$ is the number of false acceptance cases and $N_{IA}$ is the total number of impostor attempts, then FAR can be defined as: 

\begin{equation}
FAR = \biggr(\frac{N_{FA}}{N_{IA}}\biggr).
\end{equation}

Similarly, false rejection is the act of deciding that a category is an impostor while the category is actually genuine. The frequency at which false rejection are made is called false rejection rate (FRR). If $N_{FR}$ is the total number of failed rejections, and $N_{EA}$ is the total number of legitimate classification attempts, then FRR can be given as: 

\begin{equation}
FRR = \biggr(\frac{N_{FR}}{N_{EA}}\biggr).
\end{equation}

Classification accuracy, confusion matrix, FAR and FRR are evaluated in this section to evaluate the classification performance between three different types of vehicle and empty road. We have selected simple nearest neighbour classifier for this analysis. If this analysis provides acceptable results, we may use better classifier in future to improve the performance further. Table \ref{accu} shows the accuracy, confusion matrix, FRR and FAR for different of training and testing images. The following can be observed form this table.

\begin{itemize}
\item We have achieved an average of $92.6\%$ classification accuracy.
\item From the evaluated confusion matrix in the table, we can observe that, confusion between category 1 (empty road with background) and category 2 (two wheeler Honda Activa) are more. This is because the size of the two wheeler is much smaller compared to the 4 wheelers (Table \ref{car_dim}). 
\item Similarly, confusion between category 3 (Sedan car Hyundai I20) and category 4 (SUV car Ford EcoSport Titanium) are more. This is due to the structural similarity between them compared to empty road or two wheeler as is evident from Table \ref{car_dim}.
\item With less number of training images, Nearest Neighbour classifier provides $92.6\%$ classification accuracy. This may further improve by using more efficient classifier.
\end{itemize}

\begin{table*}
\caption{Vehicle Classification Performance using LTE-CommSense}
\begin{center}
\begin{tabular}{|m{25mm}|m{20mm}|m{20mm}|m{10mm} m{10mm} m{10mm} m{10mm} |m{17mm}|m{17mm}|}   \hline
Training Images & Testing Images & Accuracy (\%) & \multicolumn{4}{c|}{Confusion Matrix} & FRR & FAR  \\ \hline
200		&	250 & 94.7	& \multirow{4}{*}{}232  & 18 		& 0 	&  0 		& 0.01			& 0.07	 \\
         &		 &	    	& 					11     	& 238  	& 0 	&  1  	&          	&	    	\\ 
         &		 &	    	& 					0     	&  0 		& 240 &   10  	&         	&	    	  \\
         &		 &	    	& 					0     	&  0 		&  13 	&  237 	&         	&	    	 \\
\hline
500		&	500 & 99.5	& \multirow{4}{*}{}499  & 1 		& 0 	&  0 		& 0			& 0.002	 \\
         &		 &	    	& 					0     	& 500  	& 0 	&  0  	&          	&	    	\\ 
         &		 &	    	& 					0     	&  0 		& 500 &   0  	&         	&	    	  \\
         &		 &	    	& 					0     	&  0 		&  0 	&  500 	&         	&	    	 \\
\hline
500		& 750	 & 91.37	& \multirow{4}{*}{}667  & 82 	& 1 	&  0 		& 0.028			   & 0.1	     \\
         &		 &	    	& 					42     	& 704 & 3 	&  1 		&          		&	    		  \\ 
         &		 &	    	& 					0     	&  0 	& 700 &   50  	&         		&	    		  \\
         &		 &	    	& 	 				0     	& 18  & 80  &  670  	&         		&	    		   \\
\hline
500		& 1000 & 88.65	& \multirow{4}{*}{}828   & 166 & 4 &  2		& 0.05			& 0.17			\\
         &		 &	    	& 					85     & 899  & 8 &  8 			&          		&	    		   \\
         &		 &	    	& 					0     & 0  & 888  &   112  	&         		&	    		   \\
         &		 &	    	& 					0     & 0  & 65  &   935		&         		&	    		   \\
\hline
750		& 1000 & 93.43	& \multirow{4}{*}{}919   & 77  & 1  &  3 		& 0.02		& 0.08			\\
         &		 &	    	& 					48     & 944 & 4 &  4 			&          		&	    		  \\
         &		 &	    	& 	 				0     & 0  &  939 &   61  	&         		&	    		  \\
         &		 &	    	& 					0     & 0  & 65  &  935   	&         		&	    		   \\
\hline
1000		& 1500 & 88.95	& \multirow{4}{*}{}1451  & 49 	& 0 	 &  0 		& 0.048			& 0.03	 \\
         &		 &	    	& 					144     	& 1351  	& 1 	 &  4  	&          	&	    	\\ 
         &		 &	    	& 					0     	&  0 		& 1266 &  234  	&         	&	    	  \\
         &		 &	    	& 					0     	&  0 		&  231 &  1269 	&         	&	    	 \\
\hline
1500		& 200	 & 93.15	& \multirow{4}{*}{}1934  & 66	& 0 	&  0 		& 0.008			   & 0.033	     \\
         &		 &	    	& 					38     	& 1942 & 12 	&  8 		&          		&	    		  \\ 
         &		 &	    	& 					0     	&  0 	& 1802 &   198  	&         		&	    		  \\
         &		 &	    	& 	 				0     	& 0  & 226  &  1774  	&         		&	    		   \\
\hline
2000		& 2500 & 89.2	& \multirow{4}{*}{}2033   & 438 & 13 &  16		& 0.009			& 0.1			\\
         &		 &	    	& 					54     & 2387  & 17 & 42 			&          		&	    		   \\
         &		 &	    	& 					0     & 0  & 2042  &   458  	&         		&	    		   \\
         &		 &	    	& 					0     & 0  & 42  &   2458		&         		&	    		   \\
\hline
2500		& 3000 & 96.2	& \multirow{4}{*}{}2849   & 138  & 10  &  3 		& 0.005		& 0.05			\\
         &		 &	    	& 					51     & 2921 & 9 &  19 			&          		&	    		  \\
         &		 &	    	& 	 				0     & 0  &  2900 &   100  	&         		&	    		  \\
         &		 &	    	& 					0     & 0  & 129 &  2871   	&         		&	    		   \\
\hline
3000		& 3500 & 93.05	& \multirow{4}{*}{}3395   & 103  & 2  &  0 		& 0.04		& 0.03			\\
         &		 &	    	& 					420     & 3073 & 6 &  1 			&          		&	    		  \\
         &		 &	    	& 	 				0     & 0  &  3316 &   384 	&         		&	    		  \\
         &		 &	    	& 					0     & 0  & 57 &  3443   	&         		&	    		   \\
\hline
3500		& 4000 & 96.59	& \multirow{4}{*}{}3902 & 99  & 2  &  0 		& 0.01		& 0.02			\\
         &		 &	    	& 					187     & 3809 & 4 &  0 			&          		&	    		  \\
         &		 &	    	& 	 				0     & 0  &  3853 &   147 	&         		&	    		  \\
         &		 &	    	& 					0     & 0  & 108 &  3892   	&         		&	    		   \\
\hline
4000		& 4500 &  93.1	& \multirow{4}{*}{}4013   & 280  & 73  &  134 		& 0.004		& 0.1			\\
         &		 &	    	& 					56     & 4265 & 7 &  172 			&          		&	    		  \\
         &		 &	    	& 	 				0     & 0  &  4432 &   68 	&         		&	    		  \\
         &		 &	    	& 					0     & 0  & 452 &  4048   	&         		&	    		   \\
\hline
4000		& 5000 & 87.6	& \multirow{4}{*}{}4031  & 595  & 130  &  244 		& 0.01		& 0.32			\\
         &		 &	    	& 					97     & 4512 & 28 &  363 			&          		&	    		  \\
         &		 &	    	& 	 				0     & 0  &  4861 &   139  	&         		&	    		  \\
         &		 &	    	& 					0     & 0  & 887 &  4113  	&         		&	    		   \\
\hline
\end{tabular}
\end{center}
\label{accu}
\end{table*}

\section{Conclusion}

This work is a feasibility analysis of using LTE-CommSense for vehicle detection and subsequent classification using practical data captured with SDR platform. We observe that the CSI values corresponding to with vehicle and without vehicle cases are consistently different from each other. Scatterplot in 2D and 3D shows distinct clusters formation for with and without vehicle. Histogram of mean values of principal components from distinct region with minor overlap. Threshold based detection method was exercised for different numbers of principal components. Results shows that, using this simple technique, $90\%$ detection accuracy can be achieved. These results show that vehicle detection is feasible using LTE-CommSense with less number of principal components. 

In the subsequent classification, we have achieved $92.6\%$ classification accuracy even with less number of training images. From the confusion matrix, we can conclude that, confusion between category 1 (empty road with background) and category 2 (two wheeler Honda Activa) are more because the size of the two wheeler is much smaller compared to other vehicles considered. Likewise, confusion between category 3 (Sedan car Hyundai I20) and category 4 (SUV car Ford EcoSport Titanium) are also more because of structural similarity between them. 

This successful demonstration of vehicle detection and classification using LTE-CommSense principle with practical data promises its usability in vehicle classification, traffic condition analysis and maintenance of traffic rules.

{\textcolor{black}{In future, absolute magnitude, squared magnitude and log magnitude will be exercised in place of the mean and the detection performance will be compared. The nearest neighbour classifier will be replaced with more efficient classifiers to achieve better classification performance. Use of Doppler information to achieve higher vehicle detection and classification performance will be investigated.}}

\bibliographystyle{IEEEtran}
\bibliography{paper}

\begin{thebibliography}{10}
\providecommand{\url}[1]{#1}
\csname url@samestyle\endcsname
\providecommand{\newblock}{\relax}
\providecommand{\bibinfo}[2]{#2}
\providecommand{\BIBentrySTDinterwordspacing}{\spaceskip=0pt\relax}
\providecommand{\BIBentryALTinterwordstretchfactor}{4}
\providecommand{\BIBentryALTinterwordspacing}{\spaceskip=\fontdimen2\font plus
\BIBentryALTinterwordstretchfactor\fontdimen3\font minus
  \fontdimen4\font\relax}
\providecommand{\BIBforeignlanguage}[2]{{%
\expandafter\ifx\csname l@#1\endcsname\relax
\typeout{** WARNING: IEEEtran.bst: No hyphenation pattern has been}%
\typeout{** loaded for the language `#1'. Using the pattern for}%
\typeout{** the default language instead.}%
\else
\language=\csname l@#1\endcsname
\fi
#2}}
\providecommand{\BIBdecl}{\relax}
\BIBdecl

\bibitem{baker}
H.~Griffiths and C.~Baker, ``Passive coherent location radar systems. part 1:
  performance prediction,'' \emph{Radar, Sonar and Navigation, IEE
  Proceedings}, vol. 152, no.~3, pp. 153--159, 2005.

\bibitem{tong}
M.~Inggs and C.~Tong, ``Commensal radar using separated reference and
  surveillance channel configuration,'' \emph{Electronics letters}, vol.~48,
  no.~18, pp. 1158--1160, 2012.

\bibitem{1211012}
S.~J. Park, T.~Y. Kim, S.~M. Kang, and K.~H. Koo, ``A novel signal processing
  technique for vehicle detection radar,'' in \emph{IEEE MTT-S International
  Microwave Symposium Digest, 2003}, vol.~1, June 2003, pp. 607--610 vol.1.

\bibitem{5983805}
X.~Liu, Z.~Sun, and H.~He, ``On-road vehicle detection fusing radar and
  vision,'' in \emph{Proceedings of 2011 IEEE International Conference on
  Vehicular Electronics and Safety}, July 2011, pp. 150--154.

\bibitem{4357739}
J.~Fang, H.~Meng, H.~Zhang, and X.~Wang, ``A low-cost vehicle detection and
  classification system based on unmodulated continuous-wave radar,'' in
  \emph{2007 IEEE Intelligent Transportation Systems Conference}, Sept 2007,
  pp. 715--720.

\bibitem{7489757}
A.~Bhatta and A.~K. Mishra, ``Implementation of {GSM} channel estimation using
  open-source {SDR} environment,'' in \emph{2015 International Conference on
  Microwave, Optical and Communication Engineering (ICMOCE)}, Dec 2015, pp.
  322--325.

\bibitem{bhatta}
------, ``{GSM}-based {CommSense} system to measure and estimate environmental
  changes,'' \emph{IEEE Aerospace and Electronic Systems Magazine}, vol.~32,
  no.~2, pp. 54--67, February 2017.

\bibitem{phd_mag}
S.~Sardar, A.~K. Mishra, and M.~Z.~A. Khan, ``{LTE CommSense} for object
  detection in indoor environment,'' in \emph{IEEE Aerospace and Electronic
  Systems Magazine}, Accepted in July 2017.

\bibitem{cher}
M.~Cherniakov, R.~S. A.~R. Abdullah, P.~Jancovic, M.~Salous, and V.~Chapursky,
  ``Automatic ground target classification using forward scattering radar,''
  \emph{IEE Proceedings - Radar, Sonar and Navigation}, vol. 153, no.~5, pp.
  427--437, Oct 2006.

\bibitem{raja}
R.~R. Abdullah, A.~A. Salah, N.~H.~A. Aziz, and N.~A. Rasid, ``{V}ehicle
  {R}ecognition {A}nalysis in {LTE} {B}ased {F}orward {S}cattering {R}adar,''
  \emph{2016 {IEEE} {R}adar {C}onference ({R}adar{C}onf)}, 2016.

\bibitem{africon}
S.~Sardar, A.~K. Mishra, and M.~Z.~A. Khan, ``{LTE-CommSense} system and its
  feasibility analysis,'' \emph{$13^{th}$ edition of IEEE AFRICON 2017}, 2017.

\bibitem{akm_patent}
A.~Mishra, ``Monitoring changes in an environment by means of communication
  devices.''\hskip 1em plus 0.5em minus 0.4em\relax Google Patents, Oct.~27
  2016, { WO} Patent App. PCT/IB2016/052,235.

\bibitem{Huang2015LowPO}
T.~Huang and T.~Zhao, ``Low {PMEPR} {OFDM} radar waveform design using the
  iterative least squares algorithm,'' \emph{IEEE Signal Processing Letters},
  vol.~22, no.~11, pp. 1975--1979, Nov 2015.

\bibitem{7410692}
X.~Zhang, F.~X. Yu, R.~Guo, S.~Kumar, S.~Wang, and S.~F. Chang, ``Fast
  orthogonal projection based on kronecker product,'' in \emph{2015 IEEE
  International Conference on Computer Vision (ICCV)}, Dec 2015, pp.
  2929--2937.

\bibitem{usrp}
S.~Yun and L.~Qiu, ``Supporting wifi and {LTE} co-existence,'' in \emph{2015
  IEEE Conference on Computer Communications (INFOCOM)}, April 2015, pp.
  810--818.

\bibitem{rfx2400}
R.~{Z}itouni, S.~{A}taman, and L.~{G}eorge, ``{RF} {M}easurements of the {RFX}
  900 and {RFX} 2400 {D}aughter {B}oards with the {USRP} {N210} {D}riven by the
  {GNU} {R}adio {S}oftware,'' in \emph{2013 International Conference on
  Cyber-Enabled Distributed Computing and Knowledge Discovery}, Oct 2013, pp.
  490--494.

\bibitem{antenna}
\emph{{USRP} N200/N210 NETWORKED SERIES. $http://www.ettus.com/product/details/
  VERT2450$}.

\bibitem{openlte}
\BIBentryALTinterwordspacing
N.~{N}ikaein, R.~{K}nopp, F.~{K}altenberger, L.~{G}authier, C.~{B}onnet,
  D.~{N}ussbaum, and R.~{G}haddab, ``{D}emo: {O}pen{A}ir{I}nterface: an open
  {LTE} network in a {PC},'' in \emph{{MOBICOM} 2014, 20th {A}nnual
  {I}nternational {C}onference on {M}obile {C}omputing and {N}etworking,
  {S}eptember 7-11, 2014, {M}aui, {H}awai}, {M}aui, {UNITED} {STATES}, 09 2014.
  [Online]. Available: \url{http://www.eurecom.fr/publication/4371}
\BIBentrySTDinterwordspacing

\bibitem{1304000}
S.~Kargin, M.~Kartal, and S.~Kurnaz, ``Target detection accuracy improvement in
  synthetic aperture radar,'' in \emph{Recent Advances in Space Technologies,
  2003. RAST '03. International Conference on. Proceedings of}, Nov 2003, pp.
  675--679.

\bibitem{78300}
J.~D. Echard, ``Estimation of radar detection and false alarm probability,''
  \emph{IEEE Transactions on Aerospace and Electronic Systems}, vol.~27, no.~2,
  pp. 255--260, Mar 1991.

\bibitem{farfrr}
S.~Sardar, G.~Tewari, and K.~A. Babu, ``A hardware/software co-design model for
  face recognition using {C}ognimem neural network chip,'' in \emph{2011
  International Conference on Image Information Processing}, Nov 2011, pp.
  1--6.

\end{thebibliography}

\end{document}